\documentclass[conference]{IEEEtran}

\usepackage{blindtext, graphicx}

\def\BibTeX{{\rm B\kern-.05em{\sc i\kern-.025em b}\kern-.08em T\kern-.1667em\lower.7ex\hbox{E}\kern-.125emX}}

\usepackage{psfrag}

\usepackage[margin=0.75in]{geometry}

\usepackage{cite}

\ifCLASSINFOpdf
\else
\fi
\usepackage{amsfonts}
\usepackage{bm}
\usepackage[cmex10]{amsmath}
\usepackage{amssymb,amsthm,mathtools}
\usepackage{srcltx}

\usepackage{algorithmic}
\usepackage{algorithm}
\begin{document}
\title{Energy-Efficient Mobile-Edge Computation Offloading for Applications with Shared Data}


\author{\IEEEauthorblockN{Xiangyu He, Hong Xing, Yue Chen, Arumugam Nallanathan}
\IEEEauthorblockA{School of Electronic Engineering and Computer Science\\
Queen Mary University of London, London, UK\\
%Mile End Road, London E1 4NS\\
Email: \{xiangyu.he, h.xing, yue.chen, a.nallanathan\}@qmul.ac.uk}

}

\maketitle

\begin{abstract}
%\boldmath
Mobile-edge computation offloading (MECO) has been recognized as a promising solution to alleviate the burden of resource-limited Internet of Thing (IoT) devices by offloading computation tasks to the edge of cellular networks (also known as {\em cloudlet}). Specifically, latency-critical applications such as virtual reality (VR) and augmented reality (AR) have inherent collaborative properties since part of the input/output data are shared by different users in proximity. In this paper, we consider a multi-user fog computing system, in which multiple single-antenna mobile users running applications featuring shared data can choose between (partially) offloading their individual tasks to a nearby single-antenna cloudlet for remote execution and performing pure local computation. The mobile users' energy minimization is formulated as a convex problem, subject to the total computing latency constraint, the total energy constraints for individual data downloading, and the computing frequency constraints for local computing, for which classical Lagrangian duality can be applied to find the optimal solution. Based upon the semi-closed form solution, the shared data proves to be transmitted by only one of the mobile users instead of multiple ones. Besides, compared to those baseline algorithms without considering the shared data property or the mobile users' local computing capabilities, the proposed joint computation offloading and communications resource allocation provides significant energy saving.
\end{abstract}

\newtheorem{proposition}{\underline{Proposition}}
\newtheorem{lemma}{\underline{Lemma}}

\IEEEpeerreviewmaketitle

\section{Introduction}
%The big picture of advent of MEC (emphasize the advantage of ``edge'' rather than ``cloud"!!!)
With the advent of the era of Internet of Things (IoT), the unprecedented growth of latency-critical applications are nevertheless hardly satisfied by {\em mobile cloud computing (MCC)} alone. To cater for the low-latency requirements while alleviating the burden over backhaul networks, {\em mobile-edge computing (MEC)}, also interchangeably known as {\em fog computing} has aroused a paradigm shift by extending cloud capabilities to the very edge within the radio access network (RAN) (see \cite{mao2017survey} and the references therein).

%The general literature review of mobile edge computing in academics and industry (in past tense!!!)
%Both industry and academia have devoted constant efforts to MEC for providing the new generations mobile networks with ultra-reliable low latency communications (uRLLC). Among pioneering industrialization on fog computing, Cisco has proposed fog computing as a promising candidate for IoT architecture \cite{Bonomi2012ACM}. In academics, \cite{6574874,7264984,7442079,8269088} focused on energy efficient offloading, \cite{6678114,6787113,8269088} emphasized on cooperative multi-user cases. For example, the network energy consumption was minimized in \cite{7511367} through an iterative offloading decision algorithm. A game-theoretic approach for achieving efficient computation offloading was presented in \cite{6787113}. An offloading framework aiming for shortening the response time and reducing energy consumption was developed in \cite{6675770}. 

Both industry and academia have devoted constant effort to providing the next generation mobile networks with ultra-reliable low latency communications (uRLLC). Among pioneering industrialization on fog computing, Cisco has proposed fog computing as a promising candidate for IoT architecture \cite{Bonomi2012ACM}. In academics, \cite{6574874,7264984,7442079,6675770} focused on one-to-one offloading scheme where there is one mobile user and one corresponding cloudlet, \cite{6678114} \cite{7511367} presented multiple-user cases where there are multiple edge servers, while \cite{6787113} related to multiple-to-one scenarios where multiple mobile users offload computing to one edge server. 
%For examples, the network energy consumption was minimized in \cite{7511367} through an iterative offloading decision algorithm, a game-theoretic approach for achieving efficient computation offloading was presented in \cite{6787113}, and an offloading framework aiming for shortening the response time and reducing energy consumption was developed in \cite{6675770}. 

%The shared-data aware MEC in the emerging 5G applications helps to further improve the system performance (low latency and/or energy).
%Recently, the concept of inherent collaborative properties in terms of the offloaded data for computing was introduced in \cite{7906521}. Many mobile applications such as augmented reality (AR) and virtual reality (VR) feature this shared data property, making multiple mobile devices share parts of computing input/output in common. In \cite{8332500}, some important insights on the interplay among the social interactions within the VR mobile social network was revealed, and a significant reduce on the  end-to-end latency was achieved through stochastic optimization technique. \cite{8335683} investigated potential spatial data correlation for VR applications, in which a non-cooperative game was formulated to minimize the delay of accomplishing computation. 

Recently, the intrinsic collaborative properties of the input data for computation offloading was investigated for augmented reality (AR) in \cite{7906521}. In fact, in many mobile applications such as augmented reality (AR) and virtual reality (VR), multiple mobile devices share parts of computing input/output in common, thus making it possible for further reducing computing latency at the edge. In \cite{8332500}, some important insights on the interplay among the social interactions in the VR mobile social network was revealed, and a significant reduce on the  end-to-end latency was achieved through stochastic optimization technique. \cite{8335683} investigated potential spatial data correlation for VR applications to minimize the delay of accomplishing computation.

%The importance of joint optimization of computation offloading and communications
On another front, joint optimization of computation offloading with communications resources (such as power, bandwidth, and rate) proves to improve the performance of fog computing by explicitly taking channel conditions and communications constraints into account. In an early research \cite{4536215}, the offloading decision making was examined through the estimation of bandwidth data without considering the allocation of communication resources and channel conditions. For communications-aware computation offloading, \cite{7769867687e} minimized the local user's computation latency in a multi-user cooperative scenario, while \cite{8234686} minimized the energy consumption of remote fog computing nodes. However, these line of work have not taken the shared data feature aforementioned into account, thus failing to fully reap the advantage of fog computing.
%\cite{8269088} studies computation offloading in a multi-antenna non-orthogonal multiple access (NOMA)-based system, users partition the tasks for local processing and offloading according to the proposed algorithm to minimize the energy consumption of all users. Also in \cite{7769867687e}, the investigation of multi-user cooperative mobile-edge computing is presented, which minimizes the local user's computation latency by optimizing the task assignment jointly with the time and power allocations.

%Describe contribution of your paper in comparison the most related work, such as Simeone's student's work
In this paper, we consider a multi-user fog computing system, in which multiple single-antenna mobile users running applications featuring shared data can choose between (partially) offloading their computing tasks to a nearby single-antenna cloudlet and executing them locally, and then download the results from the cloudlet. Mobile users' overall energy consumption is minimized via joint optimization of computation offloading and communications resource allocation. Compared with existing literature, e.g., \cite{7906521}, although it investigated the energy minimization problem of shared-data featured offloading, it did not find the optimal solution. Moreover, it did not draw explicit conclusion regarding the channel condition's influence in the computation offloading. From this point of view, our work provides in-depth understanding of the shared-data featured offloading in MEC systems.

\section{System Model}
We consider a mobile-edge system that consists of $U$ mobile users running AR applications, denoted as ${\cal U}=\{1, ..., U\}$, and one base station (BS) equipped with computing facilities working as a cloudlet. All of the mobile users and the BS are assumed to be equipped with single antenna.

The input data size for user $u$ is denoted by $D^I_u$, $\forall u \in {\cal U}$, in which one fraction data size of $D_S^I$ bits are the shared data that is the same across all $U$ mobile users and the other fraction of $D_u^L$ bits are the data executed locally by user $u$. The shared data  can be transmitted from each user by part, denoted by \(D_{u,S}^I\), \(\forall u \in {\cal U}\), such that \(\sum_{u=1}^UD_{u,S}^I=D_S^I\). The amount of input data that is exclusively transmitted by $u$ is thus given by $\bar{D}^I_u=D^I_u-D^I_{S}-D^L_u, \forall u\in {\cal U}$.

\begin{figure}[h]
    \centering
    \includegraphics[width=8.5cm]{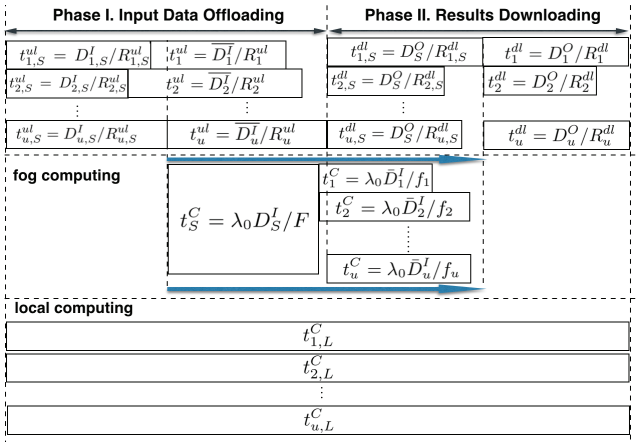}
    \caption{\bf{Timing illustration for the considered multi-user MEC system.}}
    \label{fig-sample}
\end{figure}	

It can be seen from Fig. 1 that there are two consecutive sub-phases for both input data offloading and results downloading phases: the shared and the individual data transmission. The transmission duration for offloading the shared input data is denoted by $t_{u,S}^{ul}$, $\forall u \in {\cal U}$; the offloading duration for the individual data is denoted as $t_u^{ul}$, $\forall u \in {\cal U}$; and the durations for downloading the shared and the individual output data are $t^{dl}_{u,S}$ and $t^{dl}_u$, $\forall u \in {\cal U}$ respectively. The remote computation time are also illustrated in Fig. 1, where $t_S^C$ and $t_u^C$, $\forall u \in {\cal U}$, denote that for the shared and the individual data transmitted to the cloudlet, respectively. Similarly,  $F$ and $f_u$, $\forall u \in {\cal U}$, denote the computational frequency (in cycles/s) allocated to the shared and the individual tasks, respectively, by the cloudlet. In addition, the local computation time is denoted by $t_{u,L}^C$, $\forall u \in {\cal U}$.

\subsection{Uplink Transmission}
As observed from Fig. 1, there are two consecutive uplink transmission sub-phases: the shared data and the individual data offloading \cite{7906521}. Each mobile user offloads its computation task to the cloudlet server via frequency division multiple access (FDMA). The channel coefficient from user $u$ is given by $h_u, \forall u \in {\cal U}$, which is assumed to remain unchanged during the uplink transmission duration. With the transmission power given by $p_{u,S}^{ul}$, the achievable individual data rate for offloading the shared data is expressed as:
\begin{equation}
R_{u,S}^{ul}=W_u^{ul}log_2(1+\dfrac{p_{u,S}^{ul}|h_u|^2}{N_0}), \forall u \in {\cal U},
\end{equation}
where $W_u^{ul}=\tfrac{W^{ul}}{U}$ with $W^{ul}$ denoting the overall bandwidth available for the uplink transmission, and $N_0$ is the additive white Gaussian noise (AWGN) power. Accordingly, $t_{u,S}^{ul}=D^I_{u,S}/R_{u,S}^{ul}$, and the energy consumed by the $u$-th user in the shared data offloading sub-phase is given as
\begin{equation}
E_{u,S}^{ul}=t^{ul}_{u,S}p_{u,S}^{ul}=\dfrac{t^{ul}_{u,S}}{|h_u|^2}f(\dfrac{D^I_{u,S}}{t^{ul}_{u,S}}), \forall u \in {\cal U}, \label{eq:shared data energy}
\end{equation}
where the function $f(x)$ is defined as $f(x)=N_0(2^{\tfrac{x}{W^{ul}_u}}-1)$.

Similarly, the energy consumption for the $u$-th user in the individual data offloading sub-phase is expressed as:
\begin{equation}
E_u^{ul}=t^{ul}_{u}p_{u}^{ul}=\dfrac{t^{ul}_{u}}{|h_u|^2}f(\dfrac{D^I_u-D^I_S-D^L_u}{t^{ul}_{u}}), \forall u \in {\cal U}.\label{eq:individual data energy}
\end{equation}

\subsection{Computation Model}
Based on the energy model in \cite{6787113}, given the local computing bits $D_u^L$, the energy consumption for executing local computation is given by:
\begin{equation}
E^C_u=\kappa_0\dfrac{(\lambda_0 D_u^{L})^3}{{t^{C}_{u,L}}^2}, \forall u \in {\cal U}, \label{eq:local computing energy}
\end{equation}
where $\lambda_0$ (in  cycles/bit) denotes the number of CPU cycles needed for processing one bit of input data, and $\kappa_0$ is the energy consumption capacitance coefficient.

\subsection{Downlink Transmission}
Similar to the uplink transmission, the downlink transmission phase also has two separate sub-phases: the shared and the individual results downloading. The shared output data are multicasted to the mobile users by the cloudlet at its maximum transmitting power $P_{\max}$. The achievable individual rate for the shared data downloading is thus given by
\begin{equation}
R^{dl}_{u,S}=W_u^{dl}log_2(1+\dfrac{P_{max}|g_u|^2}{N_0}), \forall u \in {\cal U},
\end{equation}
where $W_u^{dl}=\tfrac{W^{dl}}{U}$ with $W^{dl}$ denotes the overall bandwidth available for downlink transmission. The downlink channel coefficient is given by $g_u$, $\forall u \in {\cal U}$. The relation between the shared output data and the input data is given by \(D_S^O=a_0D_S^I\), where \(a_0\) is the factor representing the number of output bits for executing one bit of input data. Accordingly, $t^{dl}_{u,S}=D^O_S/R^{dl}_{u,S}, \forall u \in {\cal U}$, and thus the latency for transmitting the shared output data to all mobile users is given by
\begin{equation}
t^{dl}_S=\max_{u\in{\cal U}}\{t^{dl}_{u,S}\}.
\end{equation}
This is because the individual results downloading cannot be initiated until the shared data has finished transmission.

After the multicasting transmission, the individual output data is sent to each mobile user via FDMA. Denoting the downlink transmitting power for the $u$-th individual data by $p^{dl}_u$, the achievable rate for individual data downloading is thus expressed as:
\begin{equation}
R^{dl}_u=W^{dl}_ulog_2(1+\dfrac{p^{dl}_u|g_u|^2}{N_0}), \forall u \in {\cal U}.
\end{equation}
Similarly, denoting the individual output data size by \(D_u^O\), \(\forall u \in {\cal U}\), \(D_u^O=a_0\bar D_u^I=a_0(D_u^I-D_S^I-D_u^L)\), and \(t_u^{dl}=D^O_u/R^{dl}_u\).
%The length of receiving the output data exclusive to user $u$ is $t^{dl}_u=D^O_u/R^{dl}_u$, where $D^O_u$ is the output data size for user $u$. The relation between the individual output data and the input data for user $u$, is given by $D^O_u=a_0(D_u^I-D_S^I-D_u^L)$, where $a_0$ is the coefficient representing the number of bits returned for executing one bit input data. Besides, the shared output data size is given as $D^O_S=a_0D^I_S$.

For energy consumption,  the overall energy consumed for decoding the result sent back by the cloudlet at the $u$-th mobile user is given by \cite{7906521}
\begin{equation}
E^{dl}_u=(t^{dl}_{u,S}+t^{dl}_u)\rho^{dl}_u, \forall u \in {\cal U}, \label{eq:downloading energy}
\end{equation}
where $\rho^{dl}_u$ (in Joules/second) captures the energy expenditure per second. 

In addition, the total energy consumed by the BS for results transmission is given by,
\begin{align}
\sum_{u \in {\cal U}}\dfrac{t_u^{dl}}{|g_u|^2}f(\dfrac{a_0(D^I_u-D^I_S-D^L_u)}{t^{dl}_u}), \forall u \in {\cal U},
\end{align} which is required not to exceed $E_{\max}$ by the BS operator.

\subsection{Total Latency}
Next, we consider the overall computing latency. As illustrated in Fig. 1, it is observed the individual data downloading in Phase II cannot start until the cloudlet completes individual data computing, and the BS finishes the shared data transmission over the downlink. 
Moreover, for the individual data computing, it cannot start before either the corresponding individual data finishes offloading or the cloudlet completes the shared data computing, i.e., \(\max\{t^{ul}_{u,S}+t^{ul}_u,\displaystyle\max_{u \in {\cal U}}\{t^{ul}_{u,S}\}+t^C_S\}\). Furthermore, also seen from Fig. 1, for the shared data results, it can only start being transmitted in the downlink after the cloudlet completes the shared data computing and all the individual data finishes offloading in the uplink, i.e., \(\max\bigg\{\displaystyle\max_{u \in {\cal U}}\{t^{ul}_{u,S}\}+t^C_S, \displaystyle\max_{u \in {\cal U}}\{t^{ul}_{u,S}+t^{ul}_u\}\bigg\}\). Combining the above facts, the total computing latency is expressed as follows:
\begin{equation}
\begin{split}
\label{equ:latency expression}
&\tau_u=\max\Bigg\{\max\{t^{ul}_{u,S}+t^{ul}_u,\displaystyle\max_{u \in {\cal U}}\{t^{ul}_{u,S}\}+t^C_S\}+t^C_u,\\
&\max\bigg\{\displaystyle\max_{u \in {\cal U}}\{t^{ul}_{u,S}\}+t^C_S, \displaystyle\max_{u \in {\cal U}}\{t^{ul}_{u,S}+t^{ul}_u\}\bigg\}+t^{dl}_S\Bigg\}+t^{dl}_u ,\\
&\forall u \in {\cal U}.\\
\end{split}
\end{equation}

\section{Problem Formulation}

The overall energy consumption at the mobile users consists of three parts: data offloading over the uplink (c.f. (2) and (3)), local computing (c.f. (4)), and results retrieving (c.f. (8)), which is thus given by
\begin{equation}
\begin{split}
&E_{total}=\sum_{u\in{\cal U}}\kappa_0\dfrac{(\lambda_0D^L_u)^3}{{t^C_{u,L}}^2}+\sum_{u\in{\cal U}}\dfrac{t_{u,S}^{ul}}{|h_u|^2}f(\dfrac{D^I_{u,S}}{t^{ul}_{u,S}})\\
&+\sum_{u\in{\cal U}}\dfrac{t^{ul}_u}{|h_u|^2}f(\dfrac{D^I_u-D^I_S-D^L_u}{t^{ul}_u})+\sum_{u\in{\cal U}}(t^{dl}_{u,S}+t^{dl}_u)\rho^{dl}_u .
\end{split}
\end{equation}

The objective is to minimize the overall energy consumption given by $E_{total}$, subject to the computing latency constraints, the maximum local computing frequencies, and the total energy consumption on the individual data at the BS. Specifically, the optimization problem is formulated as below:

\begin{subequations}
\begin{align}
\mathrm{(P1)}: &{\kern-12pt}\mathop{\mathtt{min}}_{\{t_{u,S}^{ul},t^{ul}_u,t^C_{u,L},t^{dl}_u,D^L_u, D^I_{u,S}\}}{\kern-12pt}
~E_{total}\\
&{\kern20pt}\mathtt {s.t.} \notag\\
&~\tau_u \leq T_{max}, \forall u \in {\cal U},\label{eq:latency constraint}\\
&\sum_{u \in {\cal U}}\dfrac{t_u^{dl}}{|g_u|^2}f(\dfrac{a_0(D^I_u-D^I_S-D^L_u)}{t^{dl}_u}) \leq E_{max}, \label{eq:downlink energy constraint}\\
&0 \leq t^C_{u,L}\leq T_{max}, \forall u \in {\cal U},\label{eq:local latency constraint}\\
&\lambda_0D^L_u\leq t^C_{u,L}f_{u,max},\label{eq:local bits constraint}\\
&0\leq D^L_u \leq D_u^I-D_S^I, \forall u \in {\cal U},\\
&\sum_{u\in{\cal U}}D^I_{u,S}=D^I_S, D^I_{u,S}\geq 0,\label{eq:shared data constraint}\\
&t_{u,S}^{ul}\geq0, t^{ul}_u\geq0, t^C_{u,L}\geq0, t^{dl}_u\geq 0, \forall u \in {\cal U}.
\end{align}
\end{subequations}

Constraint \eqref{eq:latency constraint} and \eqref{eq:local latency constraint} gives the latency constraints that the time taken for accomplishing computing tasks cannot excess the maximum allowed length, both for offloading and local computing. \eqref{eq:downlink energy constraint} tells that the available energy for downlink transmission of remote computing node should be lower than a maximum level. \eqref{eq:local bits constraint} restricts the number of allowable local computing bits imposed by local computing capabilities. Besides, \eqref{eq:shared data constraint} puts that adding all the shared data bits offloaded by all mobile users respectively, the value should be equal to the exact amount of shared bits existing in the same user group.

\section{Optimal scheme for joint offloading and communication resource allocation}

\subsection{Problem Reformulation}

Although the latency expression \eqref{equ:latency expression} looks complex in its from, \eqref{eq:latency constraint} is still a convex constraint. For the ease of exposition, we assume herein that the cloudlet executes the shared and the individual computing within the duration of the individual data offloading and the shared results downloading, respectively, i.e., \(t_S^C\ll t_u^{ul}\), and \(t_u^c\ll t_{u,S}^{dl}\), \(\forall u \in {\cal U}\)\footnote{We assume herein that the computation capacities at the cloudlet is relatively much higher than those at the mobile users, and thus the computing time taken is much shorter than the data transmission time.}. As a result, \eqref{eq:latency constraint} can be simplified as below:
\begin{equation}
\max\{t^{ul}_{u,S}+t^{ul}_u\}+t^{dl}_S+t^{dl}_u \leq T_{max}, \forall u \in {\cal U}.\label{eq:transformed latency constraint}
\end{equation}
by introducing the auxiliary variable $t^{dl}$, which satisfies $t^{dl}_u \leq t^{dl}, \forall u \in {\cal U}$, \eqref{eq:transformed latency constraint} reduces to
\begin{equation}
t^{ul}_{u,S}+t^{ul}_u\leq T_{max}-t^{dl}_S-t^{dl}, \forall u \in {\cal U} .\label{eq:final latency constraint}
\end{equation}

Notice that \(E_u^C\)'s (c.f. (4)) is monotonically decreases with respect to the local computing time $t_{u,L}^C$ for each mobile user. To obtain the minimal energy consumption, it is obvious that $t_{u,L}^C=T_{max}, \forall u \in {\cal U}$. Then the optimization problem to be solved is reformulated as:
\begin{subequations}
\begin{align}
\mathrm{(P1^\prime)}: &{\kern-12pt}\mathop{\mathtt{min}}_{\{t_{u,S}^{ul},t^{ul}_u,t^{dl}_u,t^{dl},D^L_u, D^I_{u,S}\}}{\kern-12pt}
~E_{total}\\
&{\kern20pt}\mathtt {s.t.}\notag\\
& ~(12c-12h),  (14).\\
&t^{dl}_u \leq t^{dl}, \forall u \in {\cal U}.
\end{align}
\end{subequations}

%\begin{equation}
%\smash{\displaystyle\min_{\{{\bf t},t_{dl},D^L_u, D^I_{u,S}\}}} \sum_{u\in{\cal U}} (E^{ul}_{u,S}+E^{ul}_u+E^C_u+E^{dl}_u)
%\end{equation}
%\quad\quad s.t.
%\quad\quad\quad\quad\quad\quad(13) (14) (16) (17) (18) (19)

\subsection{Joint offloading and communication resource allocation}
Introducing dual variables ${\bm {\beta}}, {\bm {\omega}}, {\bm {\sigma}}, {\nu}$, the Lagrangian of problem $(P1^\prime)$ is presented as:
\begin{equation}
\begin{split}
&L({\bm {\beta}}, {\bm {\omega}}, {\bm {\sigma}}, {\nu},t_{u,S}^{ul},t^{ul}_u,t^{dl}_u,t^{dl},D^L_u, D^I_{u,S})=\\
&\sum_{u\in{\cal U}}\dfrac{t_{u,S}^{ul}}{|h_u|^2}f(\dfrac{D^I_{u,S}}{t^{ul}_{u,S}})+\sum_{u\in{\cal U}}\dfrac{t^{ul}_u}{|h_u|^2}f(\dfrac{D^I_u-D^I_S-D^L_u}{t^{ul}_u})\\
&+\sum_{u\in{\cal U}}\kappa_0\dfrac{(\lambda_0D^L_u)^3}{{t^C_{u,L}}^2}+\sum_{u\in{\cal U}}(t^{dl}_{u,S}+t^{dl}_u)\rho^{dl}_u+\sum_{u\in{\cal U}}\beta_u(t^{ul}_{u,S}\\
&+t^{ul}_u-T_{max}+t^{dl}_S+t^{dl})+\sum_{u\in{\cal U}}\omega_u(\lambda_0D^L_u\\
&-t^C_{u,L}f_{u,max})+\sum_{u\in{\cal U}}\sigma_u(t^{dl}_u-t^{dl})\\
&+\nu[\sum_{u\in{\cal U}}\dfrac{t^{dl}_u}{|g_u|^2}f(\dfrac{a_0(D^I_u-D^I_S-D^L_u)}{t^{dl}_u})-E_{max}],
\end{split}
\end{equation}
where ${\bm {\beta}}=\{\beta_1, ..., \beta_U\}$ are dual variables associated with the latency constraint \eqref{eq:final latency constraint}, ${\bm {\omega}}=\{\omega_1, ..., \omega_U\}$ are associated with local computing bits constraint \eqref{eq:local bits constraint}), ${\bm {\sigma}}=\{\sigma_1, ..., \sigma_U\}$ are connected with the constraint for auxiliary variable $t^{dl}$, and $\nu$ catches the downlink transmission energy constraint \eqref{eq:downlink energy constraint}. Hence, we have the Lagrangian dual function expressed as:
\begin{equation}
\begin{split}
&g({\bm {\beta}}, {\bm {\omega}},{\bm {\sigma}}, {\nu})\\
&=\smash{\displaystyle\min_{\{t_{u,S}^{ul},t^{ul}_u,t^{dl}_u,t^{dl},D^L_u, D^I_{u,S}\}}} L({\bm {\beta}}, {\bm {\omega}},{\bm {\sigma}}, {\nu},t_{u,S}^{ul},t^{ul}_u,t^{dl}_u,t^{dl},\\
&D^L_u, D^I_{u,S}),\label{eq:Lagrangian dual function}
\end{split}
\end{equation}
\quad\quad s.t.
\quad\quad\quad\quad\quad\quad (12f-12h).

Consequently, the corresponding dual problem is formulated as:
\begin{equation}
\smash{\displaystyle\max_{\{{\bm {\beta}}, {\bm {\omega}}, {\bm {\sigma}}, {\nu}\}}} g({\bm {\beta}}, {\bm {\omega}}, {\bm {\sigma}},{\nu})\label{eq:Dual Problem}
\end{equation}
\quad\quad s.t.
\begin{equation}
\quad\quad\quad\quad\quad\quad {\bm{\beta}}\succeq 0, {\bm{\omega}}\succeq 0, {\bm{\sigma}}\succeq 0, \nu\geq0. \notag
\end{equation}

\begin{proposition}
Given a determined set of dual variables \({\bm {\beta}}, {\bm {\omega}}, {\bm {\sigma}}, {\nu}\), the optimal solution to the Lagrangian dual problem (16) can be determined as follows.  \label{proposition1}

The optimal primal variables \(t_{u,S}^{ul}\), \(t_{u}^{ul}\), and \(t_u^{dl}\), are given by
\begin{equation}
\hat t^{ul}_{u,S}=\dfrac{\hat D^I_{u,S}}{\dfrac{W^{ul}_u}{ln2}[W_0(\dfrac{1}{e}(\dfrac{\beta_u|h_u|^2}{N_0}-1))+1]}, \forall u \in {\cal U}.\label{equ:share time}
\end{equation}
\begin{equation}
\label{equ:uplink time}
\hat t^{ul}_{u}=\dfrac{D^I_u-D^I_S-\hat D^L_u}{\dfrac{W^{ul}_u}{ln2}[W_0(\dfrac{1}{e}(\dfrac{\beta_u|h_u|^2}{N_0}-1))+1]}, \forall u \in {\cal U}.
\end{equation}
\begin{equation}
\label{equ:downlink time}
\hat t^{dl}_u=\dfrac{a_0(D^I_u-D^I_S-\hat D^L_u)}{\dfrac{W^{dl}_u}{a_0ln2}[W_0(\dfrac{1}{e}(\dfrac{(\rho^{dl}_u+\sigma_u)|g_u|^2}{\nu N_0}-1))+1]}, \forall u \in {\cal U}.
\end{equation}
where $W_0(x)$ is the principle branch of the Lambert $W$ function defined as the solution for $W_0(x)e^{W_0(x)}=x$ \cite{8234686}, $e$ is the base of the natural logarithm;
the optimal auxiliary variable \(t^{dl}\) is given by:
\begin{equation}
\label{equ:auxiliary tdl}
\hat t^{dl}=\left\{
\begin{aligned}
0, \quad&\sum_{u\in{\cal U}}\beta_u-\sum_{u\in{\cal U}}\sigma_u >0,\\
T_{max}-t^{dl}_S, \quad&otherwise;
\end{aligned}
\right.
\end{equation}
and the optimal local computing data size is given by
\begin{equation}
\label{equ:local bits}
\begin{split}
&\hat D^L_u=\\
&\min\Bigg\{T_{max}\sqrt{\bigg[\dfrac{N_0ln2}{3\kappa_0{\lambda_0}^3}(\dfrac{2^{\tfrac{\hat r^{ul}_{u}}{W^{ul}_u}}}{W^{ul}_u|h_u|^2}+\dfrac{\nu a_0\cdot 2^{\tfrac{a_0 \hat r^{dl}_{u}}{W^{dl}_u}}}{W^{dl}_u|g_u|^2})-\dfrac{\omega_u}{3\kappa_0 \lambda_0^2}\bigg]^+}\\
&, D^I_u-D^I_S\Bigg\}, \forall u \in {\cal U}, \notag
\end{split}
\end{equation}
where $\hat r^{ul}_{u}=\frac{W^{ul}_u}{ln2}[W_0(\frac{1}{e}(\frac{\beta_u|h_u|^2}{N_0}-1))+1]$ and $\hat r^{dl}_{u}=\frac{W^{dl}_u}{a_0ln2}[W_0(\frac{1}{e}(\frac{(\rho^{dl}_u+\sigma_u)|g_u|^2}{\nu N_0}-1))+1]$, $\forall u \in {\cal U}$.
%are the offloading and downloading transmission rates observed in \eqref{equ:uplink time} and \eqref{equ:downlink time}.
\end{proposition}

\begin{IEEEproof}
Please refer to Appendix A.
\end{IEEEproof}

In fact, on one hand, \(\hat r_u^{ul}\)'s and \(\hat r_u^{dl}\)'s can be interpreted as the optimum transmission rate for the shared/individual data offloading and the individual data downloading, respectively, given the dual variables. On the other hand, for each user $u$, the optimal transmission rate for the shared data is seen to be identical to that of the individual data over the uplink, given that the uplink channel gains remain unchanged during the whole offloading phase.

%It is noticed that obtaining the solutions of $D_u^L$ and $D_{u,S}^I$ is the prerequisite for getting the exact value of the primal variables $t^{ul}_{u,S}$, $t^{ul}_{u}$, and $t^{dl}_u$. Hence we have the expression of the local computing bits. Since $t^C_{u,L}=T_{max}$, given optimal uplink transmission data rate $\hat r^{ul}_{u}$ and optimal downlink transmission rate $\hat{r}^{dl}_u$, the expressions of the optimal value for local computing bits are given by the analysis of (24), which gives

Next, to obtain the optimal offloading bits of the shared data for each user, i.e., \(\hat D_{u,S}^I\), we need the following lemma.
\begin{lemma}
The optimal offloaded shared data for user $u$ is expressed as,
\begin{equation}
\label{equ:shared bits}
\hat D^I_{u,S}=\left\{
\begin{aligned}
D^I_S, \quad&\hat{u}=arg \smash{\displaystyle\min_{1 \leq u \leq U}} \Delta_u,\\
0, \quad&otherwise,
\end{aligned}
\right.
\end{equation}
where $\Delta_u=\frac{f(\hat r^{ul}_{u,S})}{\hat r^{ul}_{u,S}|h_u|^2}+\frac{\beta_u}{\hat r^{ul}_{u,S}}, \forall u \in {\cal U}$.
\end{lemma}

\begin{IEEEproof}
Please refer to Appendix B.
\end{IEEEproof}

Notable, it is easily observed from Lemma 1 that the shared data is optimally offloaded by one specific user instead of multiple ones.

Based on Proposition 1, the dual problem can thus be iteratively solved according to ellipsoid method (with constraints), the detail of which can be referred to \cite{EE364b}. The algorithm for solving $\mathrm{(P1^\prime)}$ is summarized in Table \ref{table: Algorithm I}.

\small\begin{table}[htp]
\begin{center}
\caption{Algorithm I for solving \(\mathrm{(P1^\prime)}\)} \label{table: Algorithm I}
\vspace{-0.75em}
\hrule
\vspace{0.50em}
\begin{algorithmic}[1]
\REQUIRE  \((\bm{\beta^{(0)}}, \bm {\omega^{(0)}}, \bm {\sigma^{(0)}}, \nu^{(0)})\)
\REPEAT
\STATE Solve \eqref{eq:Lagrangian dual function} given \((\bm {\beta^{(i)}},\bm {\omega^{(i)}},\bm {\sigma^{(i)}}, \nu^{(i)})\) according to Proposition~\ref{proposition1} and obtain \(\{\hat t_{u,S}^{ul}, \hat t_{u}^{ul}, \hat t_u^{dl}, \hat t_{dl}, \hat D_u^L, \hat D_{u,S}^I\}\);
\STATE update the subgradient of \(\bm{\beta},\bm{ \omega},\bm {\sigma}, \nu\) respectively, i.e., \(t^{ul}_{u,S}+t^{ul}_u-T_{max}+\displaystyle\max_{u \in {\cal U}}\{t^{dl}_{u,S}\}+t^{dl}\), \(\lambda_0D^L_u-t^C_{u,L}f_{u,max}\), \(t^{dl}_u-t^{dl}\), \(\sum_{u\in{\cal U}}\frac{t^{dl}_u}{|g_u|^2}f(\dfrac{a_0(D^I_u-D^I_S-D^L_u)}{t^{dl}_u})-E_{max}\) in accordance with the ellipsoid method \cite{EE364b};
\UNTIL the predefined accuracy threshold is satisfied.
\ENSURE The optimal dual variables to the dual problem \eqref{eq:Dual Problem} \((\bm{\beta^\ast},\bm{\omega^\ast},\bm{\sigma^\ast}, \nu^\ast)\)
\STATE Solve \eqref{eq:Lagrangian dual function} again with \((\bm{\beta^\ast},\bm{\omega^\ast},\bm{\sigma^\ast},\nu^\ast)\)
\ENSURE \(\{t_{u,S}^{ul\ast}, t_{u}^{ul\ast},  t_u^{dl\ast}, t^{dl\ast},  D_u^{L\ast}, D_{u,S}^{\ast}\}\)
\end{algorithmic}
\vspace{0.50em}
\hrule
\end{center}
\vspace{-1.0em}
\end{table}

\section{Numerical Results}
\begin{normalsize}
In this section, the numerical results of the proposed algorithm together with other baseline algorithms are presented. Except for the local computing only scheme where users execute all the data bits locally, there are three other offloading schemes presented as baseline algorithms: 1) {\it Offloading without considering the shared data}: the collaborative properties are ignored, every user makes the offloading decision without coordination among other users; 2) {\it Full offloading only}: the shared data is taken into consideration, but the whole chunks of input data of every user are forced to be offloaded to the edge computing node, excluding the local computing capability from participating in the computation tasks; 3) {\it Offloading with equal time length}: taking the correlated data into consideration, the data offloading and downloading are performed for each user with equal time length, with optimal solutions obtained through CVX. %The offloading without considering the shared data scheme is equivalent to solving the proposed optimization problem by forcing $D_S^I$ equal to 0, and the full offloading only scheme is equivalent to solving the proposed problem by forcing $f_{u,max}=0, \forall u \in {\cal U}$. The offloading with equal time length scheme is sovled using CVX.

In the simulation, the bandwidth avaialble is assumed to be $W^{ul}=W^{dl}$=10MHz, the maximum downlink transmit power $P_{max}=1W$, and the input data size $D_u^I=10kbits$ for all users. The spectral density of the (AWGN) power is -169 dBm/Hz. The mobile energy expenditure per second in the downlink is $\rho^{dl}_u$=0.625 J/s \cite{7906521}, the maximum local computing capability $f_{u,max}=1G$Hz. Besides, $\lambda_0=1\times 10^3$ cycle/bit, $a_0=1$, $\kappa_0=10^{-26}$. The pathloss model is $PL=128.1+37.6log_{10}(d_u)$, where $d_u$ represents the distance between user $u$ and edge computing node in kilometers.

\end{normalsize}

\begin{figure}
    \centering
    \includegraphics[scale=0.37]{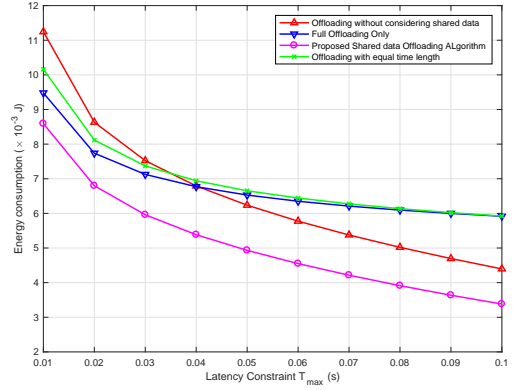}    
    \caption{\bf{Energy consumption versus different latency constraints}}
    \label{fig-sample}
\end{figure}

\begin{normalsize}
Fig.2 depicts how the energy consumption changes with different latency constraints. The energy consumption are becoming lower as the latency requirement gets longer for all listed offloading algorithms. Only the proposed offloading scheme can give the lowest energy consuming performance. The best energy saving improvement can only be achieved through the joint participation of local computing and shared data coordination. Besides, even though the equal time length offloading has lower complexity than the proposed algorithm, it cannot compete with the proposed one in terms of energy saving. Recalling our conclusion that the best way to achieve the energy saving is to let these correlated bits transmitted by one specific user, the reason is that forcing offloading time duration to be equal makes the shared data to be transmitted by all users simultaneously.

The energy consumed for computing one data bit increases exponentially as the latency constraint diminishes. Hence for the local computing only scheme, when latency constraint comes to 0.01 second the energy taken to finish the computation tasks, which is 1000 mJoules, can reach up to nearly 100 times more than those of all the offloading algorithms. Then it drops exponentially to 10 mJoules when the latency constraint goes to 0.1 second. As a result, the curve representing local computing only is not added in Fig.2, otherwise the comparison of the offloading schemes will not be clear. 
\end{normalsize}

%\begin{normalsize}
%In Fig.3, the overall energy consumption versus the different numbers of users is presented. Under the latency constraint $T_{max}=0.08s$, the energy consumption of local computing only algorithm diminishes to a relatively low level comparable with the offloading algorithms. There is no doubt that more users will incur higher overall energy consumption. It is clear that local computing only scheme induces much higher energy expenditure than any other algorithms including offloading. The proposed offloading algorithm still gives the lowest energy consumption. %Since the multiple-access scheme applied here is FDMA, which divides the available bandwidth equally to each users, more participating users means less bandwidth for each user. Then in order to maintain transmit rate in lower given bandwidth, higher power is applied. At this case, the energy need for offloading one bit data becomes higher than computing one bit data locally. So the full offloading scheme gives higher energy consuming than others when the number of users gets higher.
%\end{normalsize}

\begin{normalsize}
In Fig.3, the energy consumption changes with the percentage of shared data is demonstrated. Apparently, as long as we take the shared data into consideration when making offloading decisions, the lower overall energy consumption is achieved when the proportion of shared data gets higher. More energy will be saved when the percentage of shared data gets higher for proposed offloading scheme compared to the scheme without considering the existence of shared data. This trend applies to the full offloading only algorithm as well, because it also cares about the existence of shared data when making offloading decisions. The energy consumptions for full offloading only do not always go under that of offloading without considering shared data. That is because when given specific latency constraint, the importance of local computing capabilities diminishes in saving mobile users' energy consumption as the share of common data increases. Since most of the data will be offloaded to the edge node, few input bits would remain local for computing. Then the energy consumption of the full offloading only scenario represents that it get closer to that of the proposed algorithm when the percentage of shared data increases. Similar trend applies to the equal time length offloading as well.
\end{normalsize}

%\begin{figure}
%    \centering
%    \includegraphics[width=8.5cm]{differ_perc_new.eps}
%    \caption{\bf{Energy consumption versus different percentage of shared data}}
%    \label{fig-sample}
%\end{figure}	
\begin{figure}
    \centering
    \includegraphics[scale=0.388]{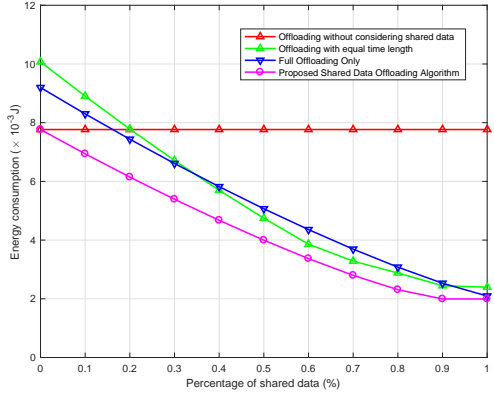}
    \caption{\bf{Energy consumption versus different percentage of shared data}}
    \label{fig-sample}
\end{figure}

\begin{equation}
t_S^C=\lambda_0 D^I_S/F
\end{equation}

\section{Conclusions}
\begin{normalsize}
In this paper, a multi-user fog computing system was considered, in which multiple single-antenna mobile users running applications featuring shared data can partially offload their individual computation tasks to a nearby single-antenna cloudlet and then download the results from it. The mobile users' energy consumption minimization problem subject to the total latency, the total downlink transmission energy and the local computing constraints was formulated as a convex problem with the optimal solution obtained by classical Lagrangian duality method. Based upon the semi-closed form solution, it was proved that the shared data is optimally transmitted by only one of the mobile users instead of multiple ones collaboratively. The proposed joint computation offloading and communications resource allocation was verified by simulations against other baseline algorithms that ignore the shared data property or the mobile users' own computing capabilities..
\end{normalsize}

\ifCLASSOPTIONcaptionsoff
  \newpage
\fi

\begin{appendices}
\section{}\label{appendix:proof of rank-one for W_p^prime}
In order to find the optimal solutions of the primary problem, we need to examine the related partial derivatives $\frac{\partial L}{\partial D^L_u}, \frac{\partial L}{\partial D^I_{u,S}}, \frac{\partial L}{\partial t^{ul}_{u,S}}, \frac{\partial L}{\partial t^{ul}_{u}}, \frac{\partial L}{\partial t^{dl}_{u}}, \frac{\partial L}{\partial t^{dl}}, \forall u \in {\cal U}$. After obtaining these partial derivatives, the KKT conditions can be applied to find the optimal solutions. For example, let $\frac{\partial L}{\partial D^L_u}$ and $\frac{\partial L}{\partial D^I_{u,S}}$ be equal to 0. The inverse function of $y=f(x)-xf'(x)$ for $x>0$ is given by $x=\tfrac{W^{ul}_u}{ln2}[W_0(-\tfrac{y}{eN_0}-\tfrac{1}{e})+1]$. Then it follows that $f(\hat r^{ul}_{u,S})-\hat r^{ul}_{u,S}f'(\hat r^{ul}_{u,S})=f(\hat r^{ul}_{u})-\hat r^{ul}_{u}f'(\hat r^{ul}_{u})=-\beta_u|h_u|^2$, and the optimal uplink transmission rate of the shared data ${\hat r^{ul}_{u,S}}$ and that of the exclusively offloaded data ${\hat r^{ul}_u}$ are thus derived. Then the expressions of the optimal primary variables are readily obtained as shown in \eqref{equ:share time}, \eqref{equ:uplink time}, \eqref{equ:downlink time}, \eqref{equ:auxiliary tdl}, \eqref{equ:local bits}, and \eqref{equ:shared bits}.

\section{}\label{appendix:proof of shared data offloading}
To obtain how the shared input data offloading $\hat D^I_{u,S}$ are distributed among users, we need to examine the partial Lagrangian regarding $D^I_{u,S}$ and $t_{u,S}^{ul}$. Replacing the shared data offloading time \(t_{u,S}^{ul}\) with \(\frac{D_{u,S}^I }{\hat r_u^{ul}}\), the partial Lagrangian is expressed as
\begin{equation}
\label{eq:partial Lagrangian}
\begin{split}
&\smash{\displaystyle\min_{\{D^I_{u,S}\}}}  \overline{L}=\sum_{u \in {\cal U}}[\dfrac{t^{ul}_{u,S}}{|h_u|^2} f(\dfrac{D^I_{u,S}}{t^{ul}_{u,S}})+\beta_u t^{ul}_{u,S}]\\
&\\
&=\sum_{u \in {\cal U}}[\dfrac{D^I_{u,S}}{\hat r^{ul}_{u,S}|h_u|^2}f(\hat r^{ul}_{u,S})+\beta_u\dfrac{D^I_{u,S}}{\hat r^{ul}_{u,S}}]\\
&=\sum_{u \in {\cal U}}\Delta_u\cdot D^I_{u,S} 
\end{split} \tag{24a} 
\end{equation}
\quad\quad s.t.
\begin{equation}
\sum_{u \in {\cal U}}D^I_{u,S}=D^I_S, D^I_{u,S}\geq 0, \forall u \in {\cal U}, \tag{24b}
\end{equation}
where we define $\Delta_u=\dfrac{f(\hat r^{ul}_{u,S})}{\hat r^{ul}_{u,S}|h_u|^2}+\dfrac{\beta_u}{\hat r^{ul}_{u,S}}$ as a constant given the dual variable \(\beta_u\)'s. As a result, the optimal solution to the linear programming (LP) (24) is easily obtained as shown in \eqref{equ:shared bits}.

\end{appendices}

\begin{IEEEbiography}[{\includegraphics[width=1in,height=1.25in,clip,keepaspectratio]{picture}}]{John Doe}
\blindtext
\end{IEEEbiography}

\bibliographystyle{ieeetr}
\bibliography{MyReference}

\end{document}